\documentclass[aps,twocolumn,floats,superscriptaddress,prd,nofootinbib]{revtex4-1}
\usepackage{graphicx, epsfig, bm, amsmath}

\usepackage{color}
\usepackage{hyperref}
\usepackage{ wasysym }
\begin{document}


\newcommand{\lcdm}{$\Lambda$CDM}

\newcommand{\gpr}{G^{\prime}}

\newcommand{\fnl}{f_{\rm NL}}
\newcommand{\curv}{{\cal R}}

\definecolor{darkgreen}{cmyk}{0.85,0.2,1.00,0.2}
\newcommand{\peter}[1]{\textcolor{red}{[{\bf PA}: #1]}}
\newcommand{\cora}[1]{\textcolor{darkgreen}{[{\bf CD}: #1]}}
\newcommand{\wh}[1]{\textcolor{blue}{[{\bf WH}: #1]}}
\newcommand{\gB}{g_B}
\newcommand{\WP}{W}
\newcommand{\XP}{X}
\newcommand{\B}{B^{\rm Bulk}}

\newcommand{\aap}{Astron. Astrophys.}


\pagestyle{plain}

\title{Fast Computation of First-Order Feature-Bispectrum Corrections}

\author{Peter Adshead}
\affiliation{Kavli Institute for Cosmological Physics,  Enrico Fermi Institute, University of Chicago, Chicago, IL 60637}
        
\author{  Wayne Hu}
\affiliation{Kavli Institute for Cosmological Physics,  Enrico Fermi Institute, University of Chicago, Chicago, IL 60637}
\affiliation{Department of Astronomy \& Astrophysics, University of Chicago, Chicago IL 60637}

\begin{abstract}
Features in the inflaton potential that are traversed in much less than an e-fold of the expansion can produce observably large non-Gaussianity.  In these models first order corrections to the curvature mode function evolution induce effects at second order in the slow roll parameters that are generically greater than $\sim 10\%$ and can reach order unity 
for order unity power spectrum features.    From a complete first order expression in 
generalized slow-roll, we devise a computationally efficient method that is as simple to evaluate as the leading order one and implements consistency relations in a controlled fashion.   This expression matches direct numerical computation  for
step potential models of
the dominant bispectrum configurations to
better than $ 1\%$ when features are small and $ 10\%$ when features are order
unity.     
\end{abstract}

\maketitle

\section{Introduction}
\label{sec:intro}

Features in the inflaton potential can give rise to  large non-Gaussianity
\cite{Chen:2006xjb,Chen:2008wn}.    In order to also satisfy cosmic microwave background (CMB)  constraints on the power spectrum,  observability in the bispectrum of an individual
feature requires that it be traversed when the horizon was of order the current horizon
in a small fraction of an e-fold \cite{Adshead:2011jq}.  
As this represents a strong violation 
of the slow-roll limit, corrections to the leading-order bispectrum expression can approach 
order unity.  
Curiously such models are in fact favored by the WMAP 7-year CMB power spectrum data
and so further tests require more accurate techniques \cite{Adshead:2011jq}. The improvement of the fit due to these models is not due to the glitch in the CMB data at $\ell = 20-40$ \cite{Peiris:2003ff, Covi:2006ci, Hamann:2007pa,Hazra:2010ve}, but comes predominantly from the region around first peak. Similar improvements have also been noted  arising from transplanckian modifications to the power spectrum
 \cite{Martin:2003sg}, 
axion monodromy inflation   \cite{Flauger:2009ab}  and non-trivial initial vacuum choices \cite{Meerburg:2011gd}. While the polarization power spectrum 
is likely to provide the next test of such models  \cite{Mortonson:2009qv,Adshead:2011jq}, the bispectrum of these models may also provide an important consistency check.

Features in the inflationary potential have a long history and have been considered by many authors going back to Starobinsky \cite{Starobinsky:1992ts} who first calculated the power spectrum due to a potential with a discontinuous first derivative. The bispectrum due to such a feature was calculated by \cite{Martin:2011sn}, see also \cite{Arroja:2011yu}. The fluctuations in the power spectrum due to a step feature in the potential were considered by \cite{Adams:2001vc, Stewart:2001cd} and their resulting non-Gaussian signatures in the bispectrum were calculated numerically by \cite{Chen:2006xjb,Chen:2008wn} and an analytic approximation was found by \cite{Adshead:2011bw, Adshead:2011jq}. Steps in the warp factor and potential in the context of brane inflation were considered by \cite{Bean:2008na}. Features in the power spectrum and large non-Gaussianities arising from resonant effects in the inflationary potential were first proposed by \cite{Chen:2008wn} and then discovered in axion monodromy inflation \cite{Flauger:2009ab, Flauger:2010ja, Leblond:2010yq}. 

An exact calculation of the bispectrum  for such models requires a computationally intensive direct integration of the curvature fluctuations for each Fourier space configuration \cite{Chen:2006xjb,Chen:2008wn}
and is impractical for analysis purposes.    In \cite{Adshead:2011bw}, a fast approximate
method was developed based on the generalized slow roll (GSR) approach.   Here the
curvature fluctuation mode functions are iteratively corrected due to the presence of the
feature.    The leading-order expression was shown to be accurate to typically ten percent for
small amplitude features and up to order unity for large amplitude features.  
In the small amplitude feature limit, this error is associated with slow-roll corrections that change the phase of the modefunctions  \cite{Adshead:2011jq}.  Even without a feature, these corrections typically generate 10\% changes to the bispectrum
\cite{Burrage:2011hd}.   A full calculation of all first-order correction terms is
computationally cumbersome  \cite{Adshead:2011bw} and again becomes impractical.

In this paper we show that the dominant first-order correction can be simply computed.
In \S \ref{sec:gsr},
we review the GSR approach and give all next to leading-order terms in the power spectrum
and bispectrum.   In \S \ref{sec:fast}, we isolate the dominant terms, test them against 
exact calculations, and establish their
compatibility with power spectrum corrections in the squeezed limit.
We discuss these results in \S \ref{sec:discussion}.

\section{GSR Approximation}
\label{sec:gsr}

In this section we review the GSR formalism \cite{Stewart:2001cd,Choe:2004zg} for
computing the power spectrum \cite{Dvorkin:2009ne} and bispectrum \cite{Adshead:2011bw} for
inflationary models with relatively large amplitude features 
including next to leading-order corrections.    Such corrections are first-order in the
GSR iteration and second order in the deviations from slow-roll.

\subsection{Power Spectrum}

Beyond the slow-roll approximation, the curvature power spectrum can be computed
exactly in linear theory as
\begin{equation}
\Delta_\curv^2 \equiv {k^3 P_\curv \over 2\pi^2} = \lim_{x \rightarrow 0} \left| { x y \over f }\right|^2 ,
\end{equation}
where
$x=k\eta$ and $\eta$ is the
conformal time to the end of inflation.  The modefunction 
 $y$ satisfies the Mukhanov-Sasaki equation, \begin{equation}
{d^2 y \over dx^2} + \left( 1 - {2 \over x^2} \right) y = {g(\ln x) \over x^2} y ,
\label{eqn:yeqn}
\end{equation}
where 
\begin{equation}
g \equiv {f'' - 3 f' \over f} ,
\end{equation}
with $' \equiv d/d\ln \eta = d/d\ln x$ and
\begin{equation}
f= 2\pi z \eta = \sqrt{ {8\pi ^2 } {\epsilon_H \over  H^2}} {a H \eta} .
\label{eqn:f}
\end{equation}
Here $\epsilon_H$ is the slow-roll parameter
\begin{equation}
\epsilon_H \equiv {1\over 2} \left(  {d \phi \over d\ln a} \right)^2,
\end{equation}
which is not necessarily small or constant.  Throughout we set the reduced Planck mass 
$M_{\rm pl}= (8\pi G)^{-1/2}=1$ as well as $\hbar=c=1$.   Note that $f''/f$ in the source function $g$ involves
first  derivatives of the second slow
roll parameter 
\begin{eqnarray}
\eta_H = -\delta_1 &\equiv& \epsilon_H - { 1\over 2} {d\ln \epsilon_H \over d\ln a}.
\label{eqn:epsilon}
\end{eqnarray}

Up to this point, no assumptions have been made beyond the validity of linear theory
but the modefunction $y$ remains an implicit functional of $\epsilon_H$.
Briefly,
the GSR approach to solving the modefunction equation (\ref{eqn:yeqn})
 is to consider the RHS as an external source with an iterative correction to $y$ \cite{Stewart:2001cd}.   
  To lowest order, we replace $y \rightarrow y_0$ where 
\begin{equation}
y_0 = \left( 1 + {i \over x} \right) e^{i x} ,
\end{equation}
is the solution to the equation with $g \rightarrow 0$.  The first-order mode function $y$
can then be obtained through  the Green function technique.   The result is an approximation that still requires $\epsilon_H$ to be small in an absolute sense but allows it to evolve rapidly with large fractional changes to its value.  
 Hence $|g|$ can be large in an absolute sense. 

To leading order, the curvature power spectrum is given by
\begin{equation}
\ln \Delta_{\curv 0}^{2} = G(\ln \eta_*) + \int_{\eta_*}^\infty {d \eta\over \eta} W(k\eta) G'(\ln \eta) ,
\end{equation}
where  $k\eta_* \ll 1$ and
\begin{equation}
W(u) = {3 \sin(2 u) \over 2 u^3} - {3 \cos (2 u) \over u^2} - {3 \sin(2 u)\over 2 u} .
\end{equation}
The source function for the power spectrum is given by
\begin{equation}
G = - 2 \ln f    + {2 \over 3} (\ln f )' ,
\end{equation}
and thus
\begin{equation}
G' = -2 (\ln f )' + {2 \over 3}  (\ln f )''  = {2 \over 3}  g - {2 \over 3} [(\ln f)']^2  .
\label{eqn:Gprime}
\end{equation}
Note that the leading order expression for the power spectrum is already first-order
in the deviations from slow roll.

The addition of the term quadratic in $(\ln f)'$ to $g$  in Eq.~(\ref{eqn:Gprime}) makes $G'$ a total derivative and
guarantees that the power spectrum
is independent of the arbitrary epoch $\eta_{*}$ after  horizon crossing,
ensuring that the curvature remains constant thereafter \cite{Dvorkin:2009ne}.  

To first order in the mode function iteration and second order in the slow-roll parameters \cite{Choe:2004zg,Dvorkin:2009ne}
\begin{equation}
 \Delta_{\curv 1}^{2} = \Delta_{\curv 0}^{2}  \left\{ [ 1+ {1\over 4}I_1^2(k) + {1\over 2}I_2(k)]^2 + {1 \over 2}I_1^2(k) \right\},
 \label{eqn:second}
\end{equation}
where
\begin{eqnarray}
I_1(k) &=& { 1\over \sqrt{2} } \int_0^\infty {d \eta \over \eta} G'(\ln \eta) X(k\eta) , \nonumber\\
I_2(k) &=& -4 \int_0^\infty { d u \over u } [ X + {1\over 3} X' ] {f' \over f} F_2(u) ,
\label{eqn:powercorrect}
\end{eqnarray}
with 
\begin{equation}
F_2(u) = \int_u^\infty {d \tilde u \over \tilde u^2} {f' \over f},
\end{equation}
and
\begin{equation}
X(u) = {3 \over u^3} (\sin u - u \cos u)^2 .
\end{equation}
When $f''/f$ controls the large deviations in $G'$, the dominant  term is
$I_1$  \cite{Dvorkin:2009ne}
\begin{equation} 
 \Delta_{\curv 1}^{2} \approx
 \Delta_{\curv 0}^{2}  \left\{ 1 + I_1^2(k) \right\}.
 \label{eqn:powerfirst}
 \end{equation}
For a wide range of models, this has been shown to be a good approximation when $|I_1 |\lesssim 1/\sqrt{2}$  \cite{Dvorkin:2011ui}.
In this approximation, the power spectrum depends only on a single source function $G'$ through
two single integrals over the functions $W$ and $X$.     We seek a similarly 
simple but accurate approximation for the bispectrum in what follows.

\subsection{Bispectrum}

For models in which large slow-roll corrections arise from a sharp feature where $\eta_H'$ or
$f''/f$ becomes large, the bispectrum can be approximated as
\cite{Adshead:2011bw}
\begin{align}\label{eqn:bispectrumgen}
B_{\curv}(k_i) &=  4 \Re\Bigg\{ i\mathcal{R}_{k_{1}}(\eta_{*})\mathcal{R}_{k_{2}}(\eta_{*})\mathcal{R}_{k_{3}}(\eta_{*})\\ 
&\quad\times \Bigg[ \int_{\eta_{*}}^{\infty} {d\eta \over \eta^2}\,  {a^{2} \epsilon_{H}}(\epsilon_{H} - \eta_{H})' (\mathcal{R}^{*}_{k_{1}}\mathcal{R}^{*}_{ k_{2}}\mathcal{R}^{*}_{ k_{3}})' \nonumber\\
&\quad +{a^{2}\epsilon_{H} \over \eta_*}(\epsilon_{H}-\eta_{H})(\mathcal{R}^{*}_{k_{1}}\mathcal{R}^{*}_{ k_{2}}\mathcal{R}^{*}_{k_{3}})' \Big|_{\eta = \eta_{*}} \Bigg]
\Bigg\} ,\nonumber
\end{align}
where the curvature fluctuation is given by
\begin{equation}
\sqrt{ {k^3 \over 2\pi^2} }  {\cal R}_k = {x y \over f},
\end{equation}
 the shorthand convention $k_i = k_1,k_2,k_3$, and $\eta_*$ is an epoch well after all modes have 
left the horizon.

Using the same iterative GSR approach to evaluate $y$, the bispectrum to first-order in the modefunction correction or second order in the slow-roll violation becomes
\cite{Adshead:2011bw} 
\begin{align}\nonumber
 B_{\curv}(k_i) 
=& {(2\pi)^4 \over 4} {\Delta_{\curv 0}(k_{1}) \over k_1^2}{ \Delta_{\curv 0}(k_{i}) \over k_2^2}{\Delta_{\curv 0}(k_{3}) \over k_3^2}\\ & \times
  \int_{\eta_*}^\infty {d\eta\over\eta}\gB(\ln\eta) [ U_0 +
U_{1A} +U_{1B}+U_{1C}
&\nonumber\\
& \qquad+U_{1D}+U_{1E}](k_i\eta) ,
\label{eqn:bispectrumfirst}
\end{align}
where the source
\begin{equation}
g_B(\ln\eta) = \frac{(\epsilon_H-\eta_H)'}{f}.
\end{equation}
The leading order response to this source comes from
\begin{equation}
U_0(k_{i}\eta)  = \left(\frac{d}{d\ln\eta}+3\right)\Re[y_{0}(k_{1}\eta)y_{0}(k_{2}\eta)y_{0}(k_{3}\eta)].
\label{eqn:zeroth}
\end{equation}
The first-order correction terms again involve $G'$ and $f'/f$
{\allowdisplaybreaks[4]
\begin{align}
 \nonumber
 U_{1A}(k_{i}\eta) =& {I_1(k_3) \over \sqrt{2}}   \left(\frac{d}{d\ln\eta}+3\right)\\ \nonumber& \times
\Im[y_{0}(k_{1}\eta)y_{0}(k_{2}\eta) [y^{*}_{0}(k_{3}\eta)+y_{0}(k_3\eta)]]\\ & \quad +{\rm cyc.}, \nonumber\\
%
%
\nonumber
U_{1B}(k_{i}\eta)=& -{1\over 2} \int_{\eta}^\infty {d\tilde{\eta} \over \tilde{\eta}} G'(\ln\tilde{\eta}) \WP(k_{3}\tilde\eta)
\\ &\times
 \left(\frac{d}{d\ln\eta}+3\right)\Re[y_{0}(k_{1}\eta)y_{0}(k_{2}\eta)y^{*}_{0}(k_{3}\eta)] \nonumber\\&\quad + {\rm cyc.},\nonumber\\
 %
\nonumber
U_{1C}(k_i\eta) = & \left[ -{I_1(k_3) \over \sqrt{2}} +
 {1\over 2} \int_{\eta}^\infty {d\tilde{\eta} \over \tilde{\eta}} G'(\ln\tilde{\eta})  \XP(k_{3}\tilde\eta)\right]\\\nonumber &\times
   \left(\frac{d}{d\ln\eta}+3\right)\Im[y_{0}(k_{1}\eta)y_{0}(k_{2}\eta)y^{*}_{0}(k_{3}\eta)]\\ & +{\rm cyc.},\nonumber\\
%
\nonumber
U_{1D}(k_i\eta) =& - {3\over 4} \int_{\eta}^\infty {d\tilde{\eta} \over \tilde{\eta}} G'(\ln\tilde{\eta}) \left({1 \over k_{3}\tilde\eta} + {1 \over (k_{3}\tilde\eta)^3}\right)\\\nonumber &
\times\left(\frac{d}{d\ln\eta}+3\right)\Im[y_{0}(k_{1}\eta)y_{0}(k_{2}\eta)\\ &
[y^{*}_{0}(k_{3}\eta)+y_{0}(k_{3}\eta)]] + {\rm cyc.},\nonumber\\
 %
 \nonumber
 U_{1E}(k_i\eta) =& - 3
\Re[y_{0}(k_{1}\eta)y_{0}(k_{2}\eta)y_{0}(k_{3}\eta)]\\ &\times {f' \over f} \left[ 1 - {1\over 2 \gB} \left( {f' \over f} \right)^2 \right],
\label{eqn:modecorrections}
\end{align} 
where cyc. denotes the 2 additional cyclic permutations of the $k$ indices.
This first-order expansion has been shown to be highly accurate even for large amplitude
features but has the drawback that it is cumbersome to compute since the nested
integrals in $U_{1B-D}$ involve configuration dependent quantities
\cite{Adshead:2011bw}.

\section{Fast Bispectrum Computation}
\label{sec:fast}

While first-order corrections to the bispectrum for models with features are generally at least
of order 10\% and hence important for accurate computation, most of the first-order terms in
Eq.~(\ref{eqn:modecorrections}) are
irrelevant where the bispectrum is observably large.    On the other hand some of these terms
are important for maintaining physicality in the superhorizon and squeezed limits. 
In \S \ref{sec:methodology} and \ref{sec:configuration}, we devise and then test a first-order  methodology that both efficiently corrects
the bispectrum where it is large and implements physicality constraints in a controlled fashion.

\begin{figure*}[t]
\centerline{\psfig{file=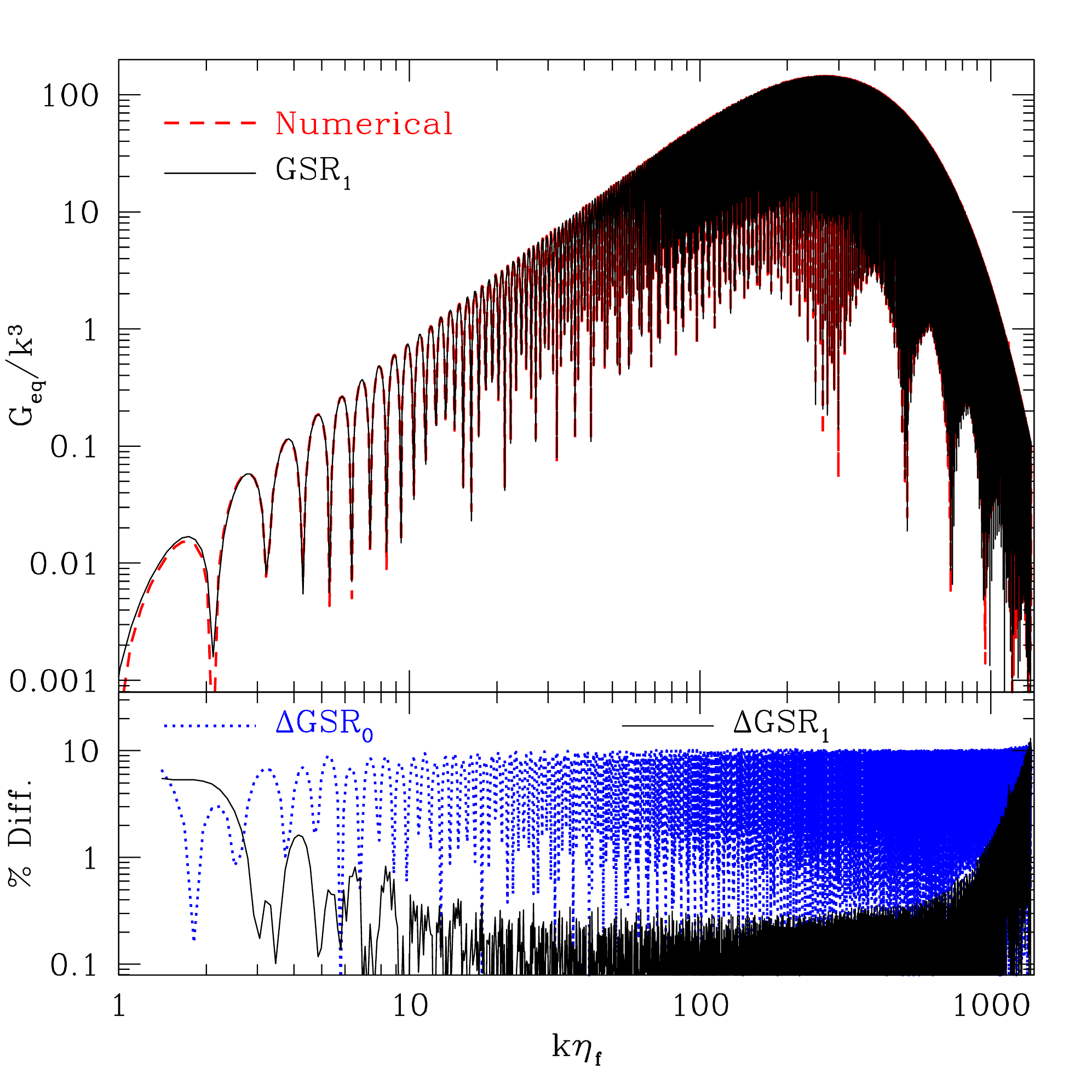,  width=3.5in} \psfig{file=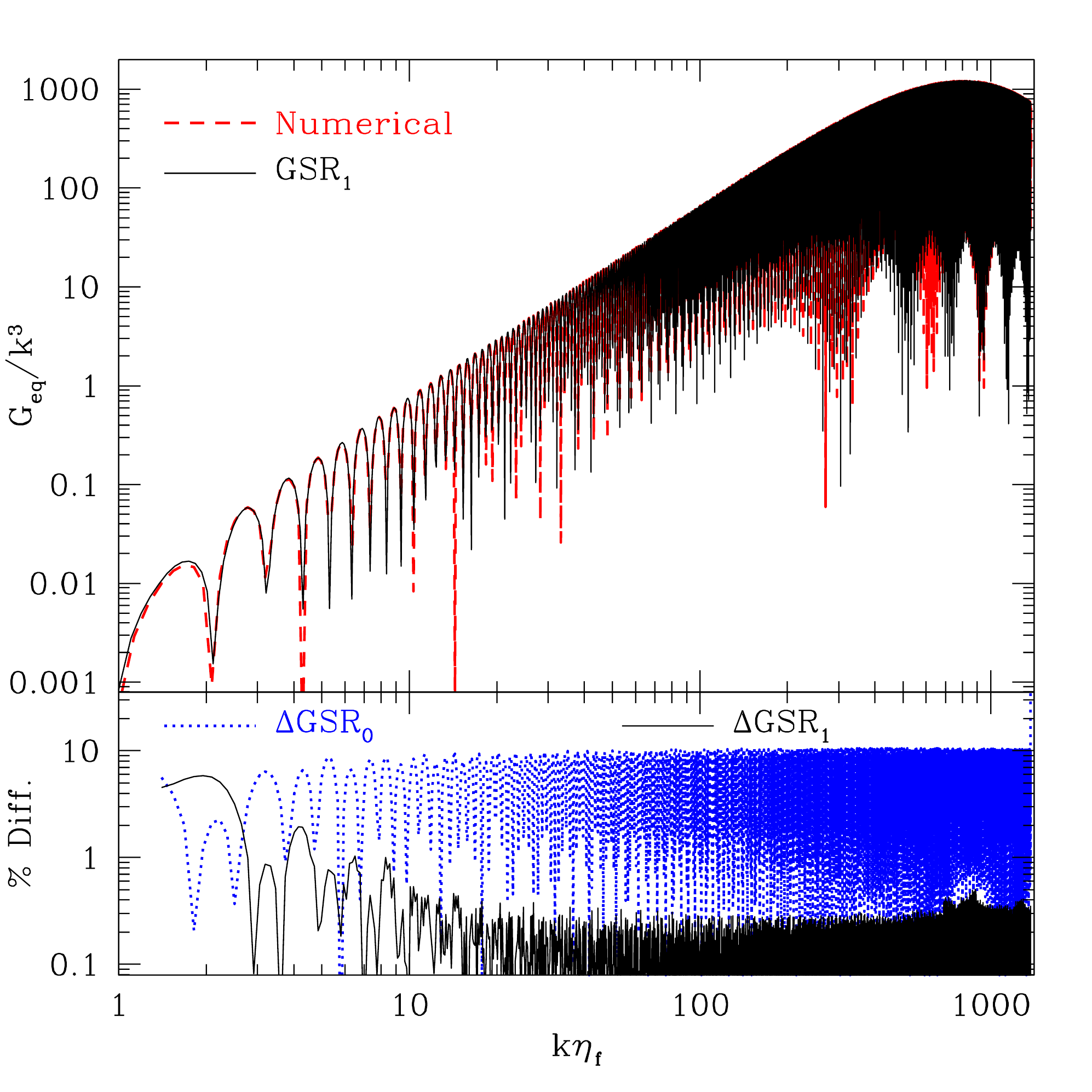, width=3.5in}}
\caption{Equilateral bispectrum for a small amplitude step $c=10^{-5}$ with $d=0.0001$ (left) and 0.0003 (right).   Upper panel shows the GSR$_1$ first order approximation versus
the full numerical computation.  Lower panel shows the percentage difference between the
two as well as between the zeroth order approximation GSR$_0$ and the numerical one.}
\label{fig:eqsmallc}
\end{figure*}

\subsection{Methodology}
\label{sec:methodology}

An observably large bispectrum arises if potential features are sufficiently sharp that
they are traversed in much less than a Hubble time.  In this case, the dominant 
contributions to the bispectrum arise when the modes are well inside the horizon $k\eta \gg 1$.     
Even if the features themselves are of small amplitude, first-order corrections from the
slow-roll contributions to $G'$ make substantial fractional corrections to the
bispectrum \cite{Adshead:2011jq}.    In this case $G'$ is
nearly constant and we can evaluate the expressions in Eq.~(\ref{eqn:modecorrections}).
We derive our method from these two assumptions but show that the resulting expressions
remain a good approximation even when features in the power spectrum reach order unity or
when one of the sides of the triangle is outside the horizon.

Under these two assumptions,
 all the nested integrals contribute very little to the integrand in Eq.~(\ref{eqn:bispectrumfirst}) in that most of their impact is around or after horizon crossing rather than before.   
In addition, models with large bispectra contribute mainly through
the $f''/f$ terms in $g_B$ and $G'$ and so $U_{1E}$ is negligible.
 The remaining
terms are the ones proportional to $I_1(k)$ in $U_{1A}$ and $U_{1C}$.   In fact the
$y_0 y_0 y_0^*$ contributions of these two terms cancel exactly leaving only $y_0 y_0 y_0$ contribution
similar to the zeroth order term (\ref{eqn:zeroth}).  As noted in \cite{Adshead:2011jq}, this fact means that for 
 triangles where all three $k\eta \gg 1$, the first-order corrections can be cast
in the same form as that of the zeroth order term, namely in terms of single integrals
involving only the perimeter of the triangle $K=k_1+k_2+k_3$.

As $k\eta$ decreases, the $I_1$ and integral contributions to $U_{1C}$ cancel, leaving
the $U_{1A}$ term as the dominant correction.  However as $k\eta \rightarrow 0$, the other 
first-order corrections ensure that the bispectra remain constant in accord with conservation of
the comoving curvature ${\cal R}_k$.   This condition can be alternately enforced by
replacing the bispectrum source with a total derivative
\cite{Adshead:2011bw}
\begin{equation}
g_B \rightarrow G_B'(\ln\eta) = \left( \frac{\epsilon_H-\eta_H}{f} \right)',
\end{equation}
just as in the case of the power spectrum.
Combining these considerations, our fast first-order bispectrum expression becomes
\begin{align}
\label{eqn:gsrlbi}
 B_{\cal R}(k_i)  & \approx  {(2 \pi)^4  \over k_1^3 k_2^3 k_3^3} {\Delta_{{\cal R}0}(k_1) 
\Delta_{{\cal R}0}(k_2)
\Delta_{{\cal R}0}(k_3)\over 4}\nonumber\\&\quad
\Big[ -I_{B0}(K) k_1 k_2 k_3 - I_{B1}(K) \sum_{i \ne j} k_i^2 k_j 
\nonumber\\&\quad  
+ I_{B2}(K) K \sum_i  k_i^2+\sum_m {I_1(k_m) \over \sqrt 2} 
\nonumber\\&\quad  \times
 \Big( -I_{B3}(K) k_1 k_2 k_3 
 - I_{B4}(K) \sum_{i \ne j} k_i^2 k_j 
 \nonumber\\&\quad
+ I_{B5}(K) K \sum_i k_i^2   
\\
& \quad
+  I_{B4}(K-2 k_m)   \sum_{i\ne j} k_i^2 k_j( 1- 2 \delta_{m j})
\nonumber\\
& \quad
- (K-2k_m)I_{B5}(K-2 k_m)  \sum_i k_i^2
\Big) \Big], \nonumber
\end{align}
where 
\begin{equation}
I_{Bn}(K) = G_B(\ln\eta_*) W_{Bn}(K\eta_*)
+   \int_{\eta_*}^\infty {d\eta \over \eta} G_B' W_{Bn}(K\eta)
\end{equation}
and
\begin{alignat*}{3}
 W_{B0}(x) & = x \sin x, 
&\qquad
 W_{B3}(x) & = -x\cos x ,
\nonumber\\
 W_{B1}(x) & = \cos x,  
&\qquad 
 W_{B4}(x) & = \sin x , 
\nonumber\\
 W_{B2}(x) &={\sin x / x},
&\qquad 
 W_{B5}(x)& = -{\cos x / x}.
\end{alignat*}
The lack of a $K-2k_k$ term in $I_{B3}$ comes from the $x\gg 1$ 
cancellation of the $U_{1A}$ and $U_{1C}$ terms.  

We shall call the $n=0-2$ terms the zeroth order approximation and the full set
$n=0-5$ the first order approximation, denoting these GSR$_{0}$ and GSR$_{1}$ respectively.  Note that the boundary term at $K\eta_* \ll 1$ is formally only defined correctly for the
$n=1,2$ terms where $W_{Bn} \rightarrow 1$ but we retain the others for notational compactness as their contributions vanish in this limit.
While the $I_{B5}$ term diverges as $K\eta_* \rightarrow 0$ we shall see that
the differencing construction in Eq.~(\ref{eqn:gsrlbi}) guarantees that this divergence or the corresponding dependence on the arbitrary scale $\eta_*$ has no
observable consequence.

The computational cost of GSR$_1$ is only double that of GSR$_0$ involving  6 rather than 3  one dimensional integrals in addition to the power spectrum
correction $I_1(k)$ from Eq.~(\ref{eqn:powercorrect}).  That the corrections are all proportional to  $I_1(k)$ is an important feature in our construction and we shall see 
enforces compatibility between the power spectrum and bispectrum approximations.

Although this construction is motivated by the slow-roll corrections to the bispectrum, we shall
see next that it also works quite well for case where the correction is dominated by the feature itself.

\begin{figure*}[t]
\centerline{\psfig{file=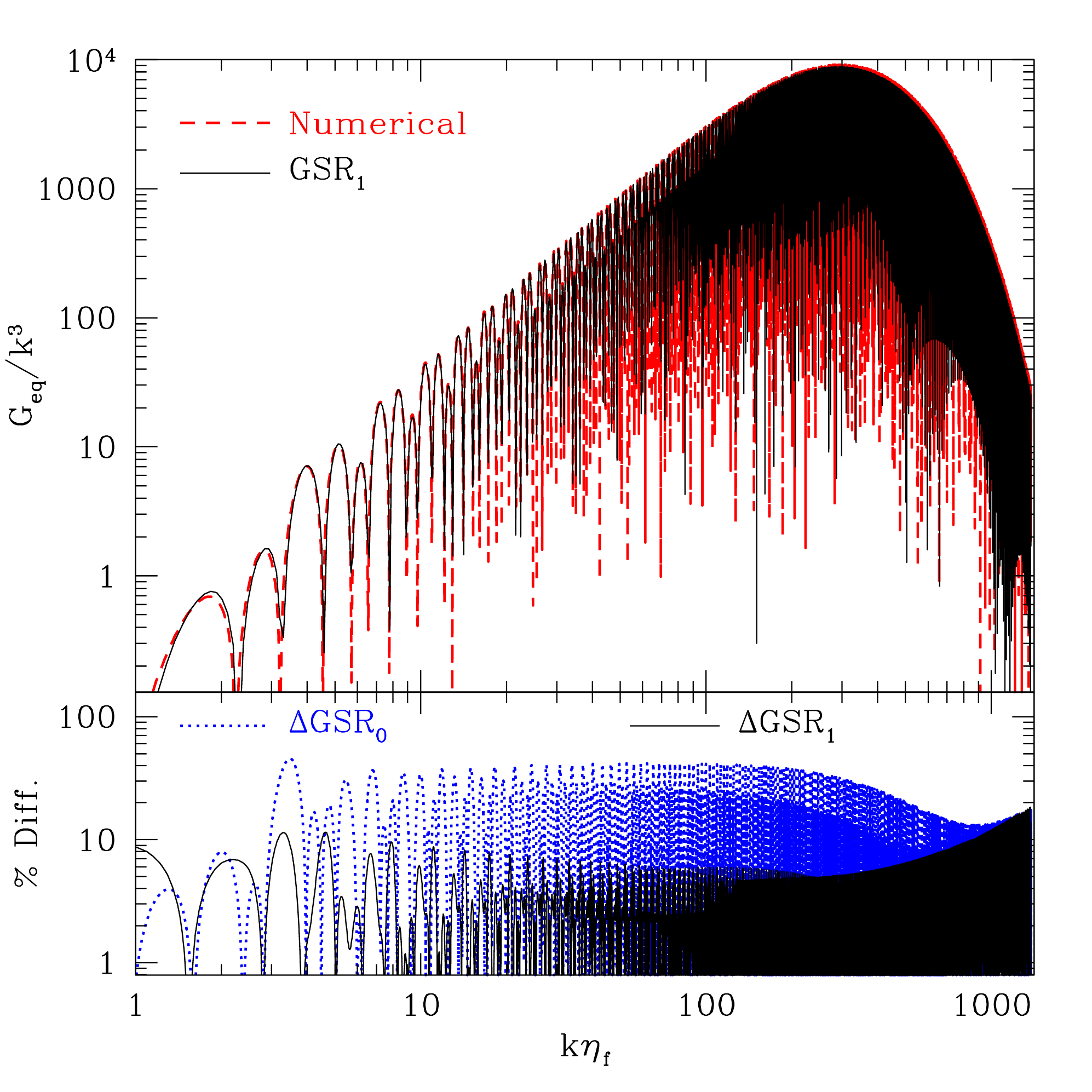, width=3.5in} \psfig{file=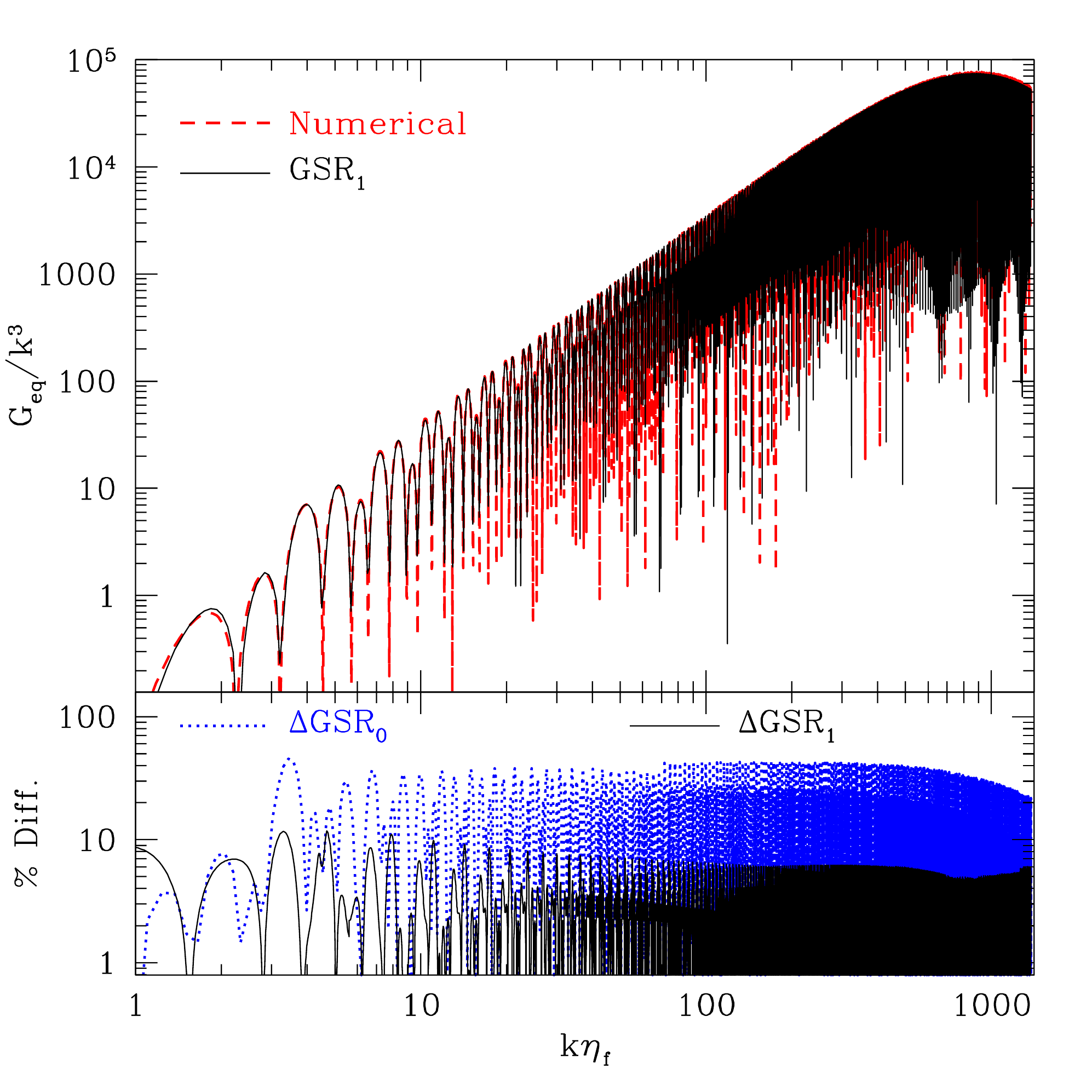, width=3.5in}}
\caption{Equilateral bispectrum for a large amplitude step $c=5.75 \times 10^{-4}$ with $d=0.0003$ (left) and 0.0001 (right).  
 Upper panels shows the GSR$_1$ first order approximation versus
the full numerical computation.  Lower panels shows the percentage difference between the
two as well as between the zeroth order approximation GSR$_0$ and the numerical one.
}
\label{fig:eqlargec}
\end{figure*}

\subsection{Configuration Tests}
\label{sec:configuration}

As an example of a feature model with a possibly observable bispectrum, we consider the  potential \cite{Adams:2001vc}
\begin{align}
V(\phi) = \frac{1}{2}m^{2}\phi^{2}\left[1+c\tanh\left(\frac{\phi - \phi_{s}}{d}\right)\right],
\end{align} 
which corresponds to a smooth step at $\phi = \phi_{s}$ of fractional height $c$ and width $d$.
Given the WMAP7 preference for a large scale feature, we set $\phi_s$ to reproduce
the scale of the maximum likelihood (ML) solution
 $\eta(\phi_s)=8.163$ Gpc \cite{Adshead:2011jq}.   For the amplitude, we take the case
 where  $c \rightarrow 0$ and the maximum likelihood solution for the WMAP7 data $c= 5.75 \times 10^{-4}$.    For the width $d$ we take several cases that would lead to observable bispectra with this amplitude and position. 
We choose $m$ to be compatible with the WMAP7 slope for the underlying slow-roll potential $m= 7.126 \times 10^{-6}$ such that the underlying tilt from the smooth part of the potential is $\bar n_s \approx 0.963$.
Following the notation of the existing literature \cite{Chen:2006xjb,Chen:2008wn}, we construct the bispectrum statistic  
\begin{align}
\mathcal{G}(k_{1}, k_{2},k_{3}) &= \frac{k_{1}^{3}k_{2}^{3}k_{3}^{3}}{(2\pi)^{4}\tilde{A}_{S}^2}B_{\curv}(k_1,k_2,k_3),
\label{eqn:dimensionlessbi}
\end{align}
where we take $\tilde{A}_{S}=2.39\times10^{-9}$ which is approximately the amplitude of the power spectrum in the
absence of the feature.

Most of the observable impact of features in the bispectrum comes from equilateral triangles where
$k_1 \sim k_2 \sim k_3$  \cite{Adshead:2011jq}.  In this limit
\begin{eqnarray}
&& B_{\cal R}(k,k,k)  \approx  {(2 \pi)^4  \over k^6 } {\Delta_{\cal R}^3(k) 
\over 4}\\&&\quad 
\times \Big\{- I_{B0}(3k) - 6 I_{B1}(3k) + 9 I_{B2}(3k) 
\nonumber\\
&&\qquad    +{ 3I_1(k) \over \sqrt 2}  \Big[ - I_{B3}(3k) - 6 I_{B4}(3k) + 9 I_{B5}(3k) 
\nonumber\\&&\qquad 
+ 2 I_{B4}(k) - 3 I_{B5}(k) \Big]\Big\} .\nonumber
\label{eqn:equilateral}
\end{eqnarray}
Note that
\begin{equation}
3 I_{B5}(3k)-  I_{B5}(k) = \int_{\eta_*}^\infty {d\eta \over \eta} G_B' \left[ - {\cos(3k\eta)-\cos(k\eta)\over k\eta}\right]
\end{equation}
and so  as $k\eta \rightarrow 0$ the quantity in brackets vanishes and the
expression becomes independent of the arbitrary end point $\eta_*$.

In Fig.~\ref{fig:eqsmallc} (upper), we compare the first order approximation (GSR$_1$) to the full numerical
calculation for the equilateral case and a small amplitude step $c=10^{-5}$ with $d=0.0001$ and $0.0003$.
Differences in the upper panel are quoted as percentages of a smooth envelope
\begin{equation}
\frac{27}{4} \frac{c}{(\epsilon_0+3c)} \frac{\Delta^{3}_{\mathcal{R}0}(k)}{\tilde A_{S}^{3/2}}(k\eta_f)^2
\frac{k/k_D}{\sinh(k/k_D)},
\label{eqn:bienvelope}
\end{equation}
 in the lower panel
to avoid dividing by an oscillatory quantity.
Here the equilateral damping scale
\begin{align}
k_D = \frac{2}{3} \frac{ \sqrt{2\epsilon_0 + 6c} }{\pi d \eta_f},
\end{align}
and $\epsilon_0 = 0.00925$ is the value calculated on the slow-roll background in the absence of the step. 
 The form of the envelope is derived from the analytic solutions in \cite{Adshead:2011jq} and is accurate at the several percent level
for all models considered here.

For $c=10^{-5}$,
the agreement of GSR$_1$ and the exact numerical treatment is excellent, with differences below the 1\% level between the first  oscillation and the damping scale
set by the finite width $d$ (upper panel).   The first order correction eliminates the 10\% errors 
of the GSR$_0$ approximation that appear mainly as a phase error (lower panel). 
Deep in the damping tail, the first order solution develops a small
phase difference leading to larger fractional errors but controlled amplitude
errors. This error can be attributed to terms in the first order expansion at Eqn.\ (\ref{eqn:modecorrections}) that are not captured by the approximation at Eqn.\ (\ref{eqn:gsrlbi}). In particular, contributions due to some of the nested integrals are damped more slowly than the leading order contributions. This means that deep in the damping tail they can become a significant fraction of the leading order result. However, since this effect only becomes important in the region where the bispectrum is already small, we conclude that it can be safely ignored.

This good agreement persists until the feature makes order unity changes to the power
spectrum.    In Fig.~\ref{fig:eqlargec}, we show the maximum likelihood amplitude model.
Here the oscillations take on a modulated form where every third extrema is reduced
in amplitude reflecting the strong oscillations in the power spectrum.   These are well 
captured by the first order approximation between the first oscillation and the damping
tail with residuals around $\sim 6\%$ correcting the $\sim 40\%$ errors of the zeroth order
approximation.   Errors in the damping tail grow again mainly due to a phase error but remain small until the bispectrum amplitude has damped to an unimportant level.
Note that in this example the maximum of $|I_1|\approx 0.25$.  Like the power spectrum, this quantity monitors the accuracy of the first order computation.   The criteria for equilateral bispectrum
is slightly more stringent than the power spectrum due to the 3 $k$-modes that can be corrected and hence
\begin{equation}
|I_1| \lesssim \frac{1}{3\sqrt{2}}
\end{equation}
 is the rough criteria for better than 10\%  percent level accuracy.

\begin{figure}[t]
\psfig{file=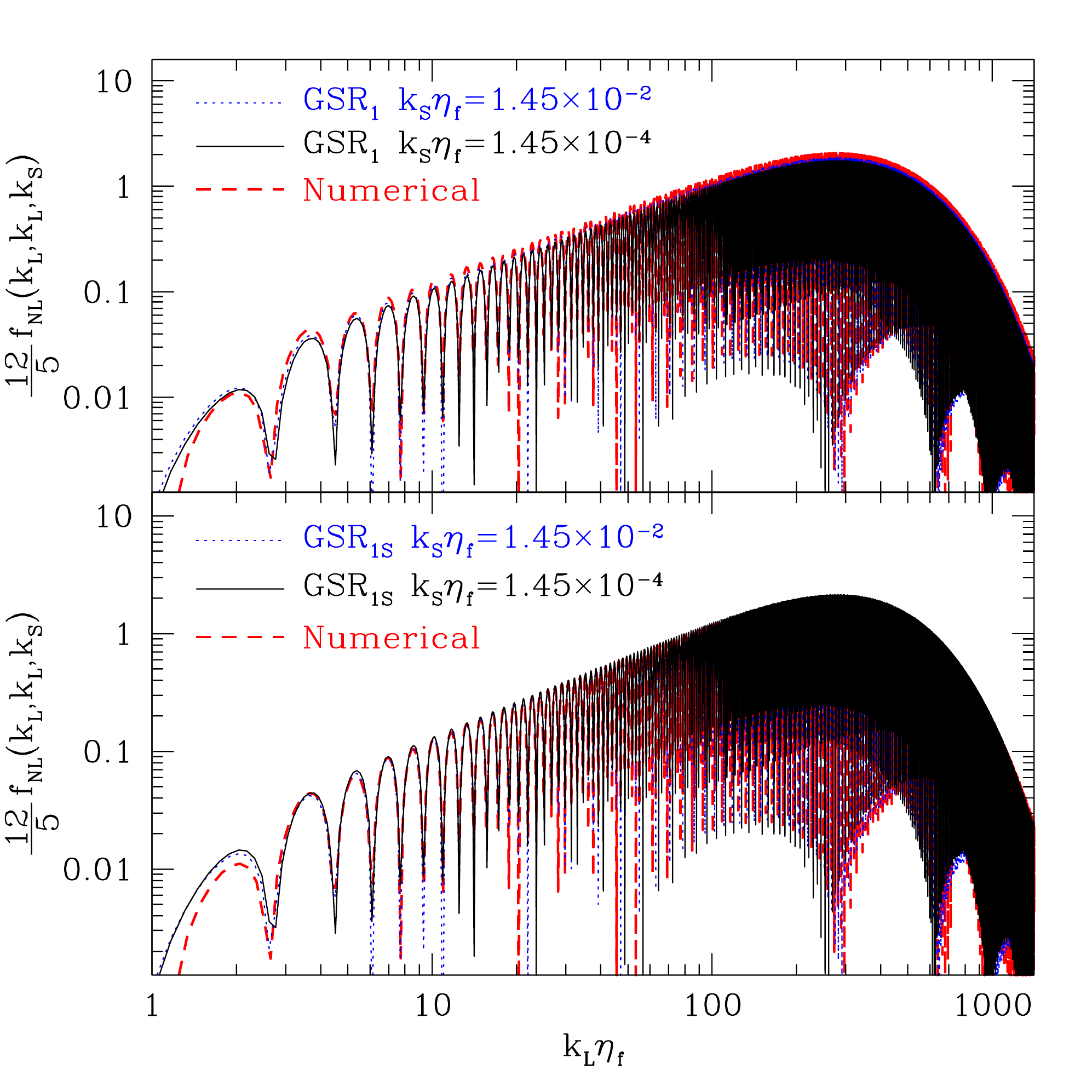, width=3.5in}
\caption{Squeezed limit bispectrum for a step with small amplitude $c = 10^{-5}$, $d=0.0003$.  Shown here is $12f_{\rm NL}(k_L, k_L, k_S)/5$ for various values of the short side, $k_S$. In the upper panel, we show the result of evaluating the first order approximation
GSR$_{1}$ versus the full numerical computation.  The lower panels here show the slow roll
$k_S$ rescaling GSR$_{\rm 1S}$ from Eq.\ (\ref{eqn:SLSRcorrection}).}
\label{fig:sqlimitlowc}
\end{figure}

\begin{figure}[t]
\centerline{\psfig{file=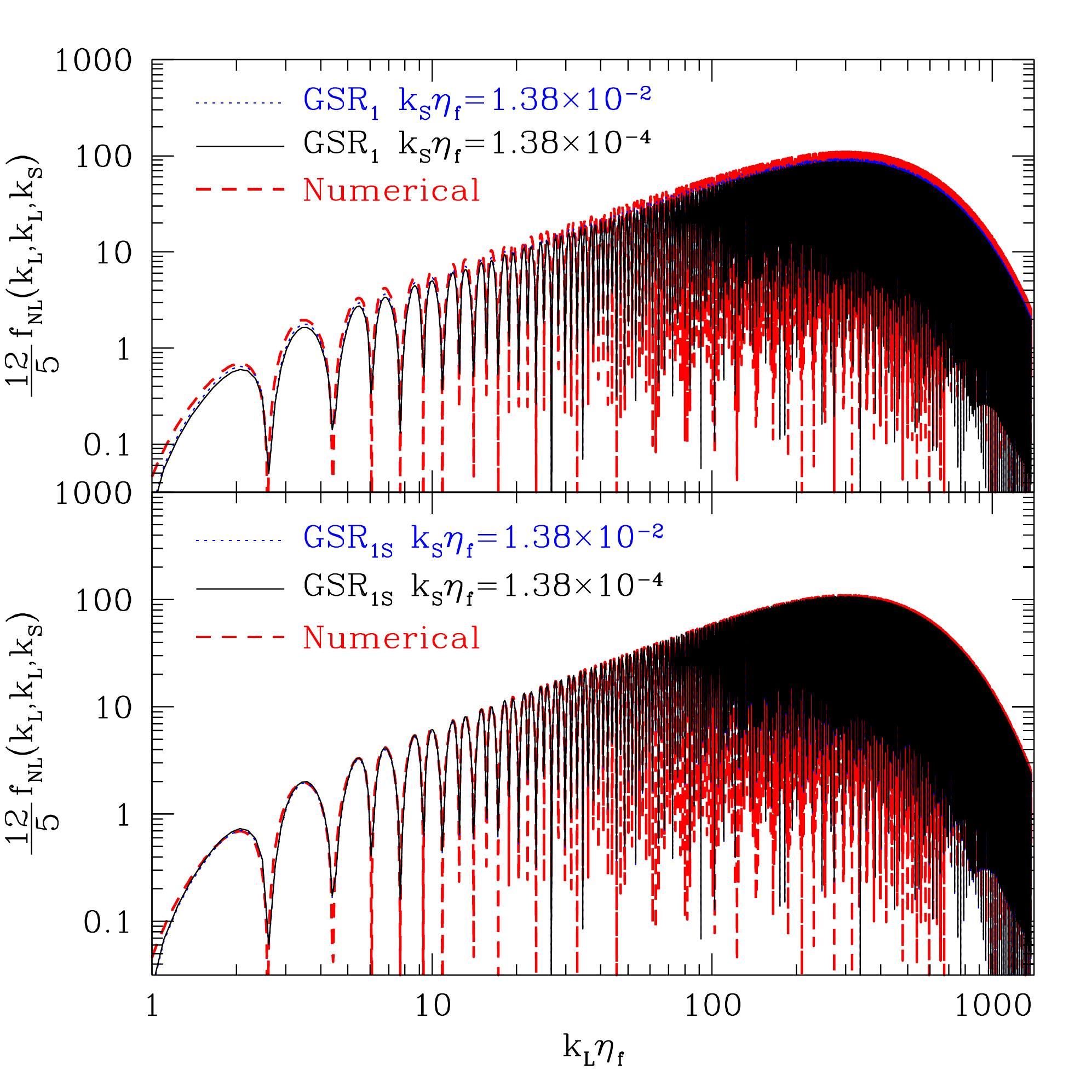, width=3.5in}}
\caption{Squeezed limit bispectrum for a  step with large amplitude $c=5.75 \times 10^{-4}$ and $d=0.0003$.  Shown here is  the GSR$_1$ and slow-roll corrected GSR$_{\rm 1S}$
approximations compared with the numerical computation as in Fig.~\ref{fig:sqlimitlowc}.   Note that the latter corrects all of the $k_S$
dependent error.}
\label{fig:sqlimithighc}
\end{figure}

An important check of the physicality of the bispectrum is the squeezed limit where
$k_1=k_S \ll k_2 \approx k_3 = k_L$ and the the bispectrum must satisfy consistency
with the power spectrum \cite{Maldacena:2002vr}
\begin{align}
n_s(k_L)-1\equiv {d \ln \Delta^2_{\cal R} \over d\ln k} \Big|_{k_L}  &= - {B_{\cal R}(k_S,k_L,k_L) \over P_{\cal R}(k_S) P_{\cal R}(k_L) } \nonumber\\  
& \equiv -\frac{12}{5} f_{\rm NL} (k_S,k_L,k_L).
\end{align}
Note that this is a non-trivial check on our construction even for small amplitude features
since we retained only the leading order corrections when all three modes are subhorizon scale during feature crossing.   

For squeezed configurations, our first-order approximation 
(\ref{eqn:gsrlbi})  becomes
\begin{eqnarray}
&& B_{\cal R}(k_S,k_L,k_L)  \approx  {(2 \pi)^4  \over k_L^3 k_S^3 } {\Delta_{\cal R}(k_S) 
\Delta_{\cal R}^2(k_L)\over 4}\\&&\quad  \times
\Big\{  - 2 I_{B1}(2k_L) 
+ 4I_{B2}(2k_L) + {4 I_1(k_L) \over \sqrt 2} 
\nonumber\\
&& \qquad \times \Big[   - I_{B4}(2k_L)
+2 I_{B5}(2k_L)
- {k_S \over k_L}  I_{B5}(k_S) \Big]
\Big\}. \nonumber
\label{eqn:squeezed}
\end{eqnarray}
Since
\begin{eqnarray}
&&2 I_{B5}(2k_L) - {k_S \over k_L} I_{B5}(k_S)\nonumber \\&&\qquad
= \int_{\eta_*}^\infty { d \eta \over \eta} G_B' \left[ - {\cos(2k_L\eta) \over  k_L\eta} + 
{\cos(k_S\eta) \over k_L \eta} 
\right] \nonumber\\
&&\qquad\approx 
\int_{\eta_*}^\infty { d \eta \over \eta} G_B' \left[ - {\cos(2k_L\eta)-1 \over  k_L\eta} \right],
\end{eqnarray}
the expression becomes
independent on the arbitrary end point $\eta_*$.

To check the consistency relation, it is useful  to  combine all of the terms to form a single integral over the source
\begin{equation}
 B_{\cal R}(k_S,k_L,k_L)  \approx  {(2 \pi)^4  \over k_L^3 k_S^3 } {\Delta_{\cal R}(k_S) 
\Delta_{\cal R}^2(k_L)\over 4} I_{\rm sq}(k_L), 
\end{equation}
where 
\begin{equation}
I_{\rm sq}(k) = G_B(\ln\eta_*) W_{\rm sq}(k\eta_*)
+   \int_{\eta_*}^\infty {d\eta \over \eta} G_B' W_{\rm sq}(k\eta),
\end{equation}
and
\begin{eqnarray} 
W_{\rm sq}(x)  &=& -2 \left[ \cos(2x) - {\sin (2 x) \over x} \right] \nonumber\\
&& - {4 I_1 \over \sqrt{2}} \left[ \sin(2 x) + {\cos (2 x) -1 \over x} \right],
\end{eqnarray}
with $I_1=I_1(k_L)$.
Using the approximation of Eq.~(\ref{eqn:powerfirst}) in the $|I_1| \ll 1$ limit
\begin{equation}
{d \ln \Delta^2_{\cal R} \over d\ln k}  \approx \int d\ln \eta W_{P}'(k\eta) G'(\ln\eta) ,
\end{equation}
where
\begin{equation}
W_{P}(x) = W(x) + {2 I_1\over \sqrt{2}} X(x) .
\end{equation}

To establish the consistency relation, we need to relate $W_{\rm sq}$ to $W_{P}$ and $G_B'$ to  $G'$.  
We can manipulate the latter pair via integration by parts
\begin{eqnarray}
 && \int_{\eta_*}^\infty d\ln  \eta W_{P}'(x) G'(\ln\eta) \nonumber\\
 && \quad = 
  -2 \int_{\eta_*}^\infty d\ln  \eta  W_{P}'(x) \left[  {f' \over f} - {1 \over 3} \left( {f' \over f} \right)' \right] \nonumber\\
 && \quad = -2  \int_{\eta_*}^\infty d\ln  \eta    {f' \over f}  [W_{P}' + {1\over 3} W_{P}''] \nonumber\\
 &&\quad =  2 {f' \over f}  [W_{P} + {1\over 3} W_{P}'](x_*)  \nonumber\\
 &&\qquad + 2  \int_{\eta_*}^\infty d\ln  \eta \left( {f' \over f} \right)' [ W_{P} + {1 \over 3} W_{P}' ] \,.
 \end{eqnarray} 
 For bispectra that are dominated by $f''/f$,
  $G_B \approx -(f'/f^2)$.  If we further take the approximation that $f \approx \Delta_{\cal R}^{-1} \approx $ const., the consistency relation is satisfied since
  \begin{equation}
  2W_{P} + {2 \over 3} W_{P}' = W_{\rm sq}.
  \end{equation}
   Thus the consistency relation holds in our first order bispectrum approximation  as long as $f$ remains nearly constant. 
 
 Of course $f$ remains constant only at zeroth order in slow roll so there are additional correction in a full first order calculation.  
 In Fig. \ref{fig:sqlimitlowc} (upper), we show that correspondingly 
 there is a small amplitude mismatch between the numerical and the GSR$_1$ results which grows logarithmically with decreasing $k_S$.  Recall that the quantity   $f_{\rm NL}(k_S, k_L, k_L)$ should become independent of $k_S$ as $k_S \to 0$ to satisfy the consistency relation.
 
 \begin{figure}[t]
\centerline{\psfig{file=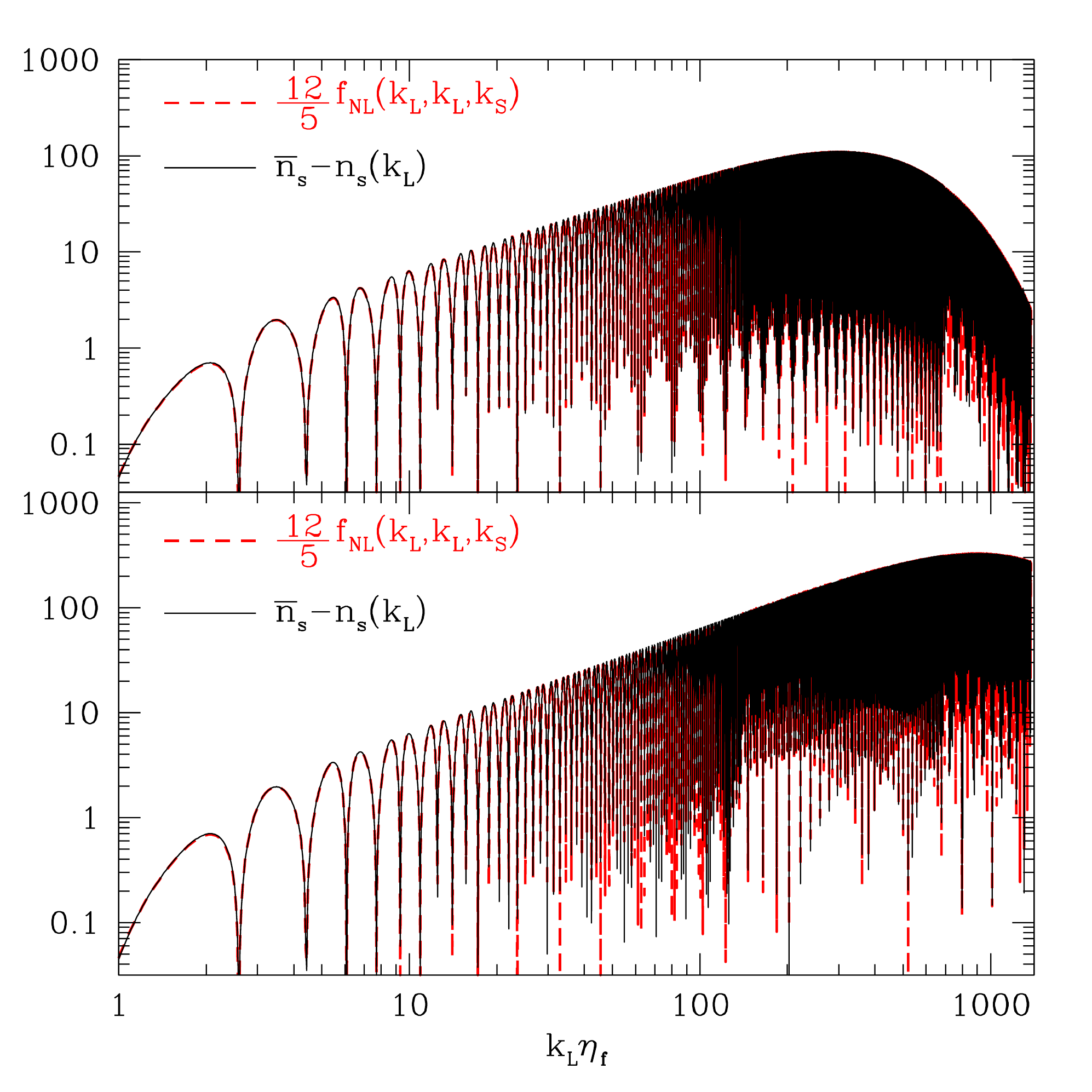, width=3.5in}}
\caption{Consistency relation test for a large amplitude step $c=5.75 \times 10^{-4}$ and $d=0.0003$ (upper) and $0.0001$ (lower).   Here the change in the power spectrum slope due to the feature $\bar n_s - n_s$ and
the squeezed bispectrum from the feature $f_{\rm NL}$ are both evaluated numerically.  For the latter we take
$k_S\eta_f=1.38\times10^{-2}$, though the results are independent of this choice.
}
\label{fig:sqlargec}
\end{figure}

One can gain further insight into the missing term  by re-examining the slow-roll limit of the full first-order expression in
 Eq.~(\ref{eqn:modecorrections}) beyond the subhorizon approximation.  Here $G' \approx 1-\bar n_s$, where recall $\bar n_s$ is the tilt that describes the power spectrum in the absence of the slow-roll violating feature.   
 The leading contributions are from the $U_{1B}$ and $U_{1D}$ terms and account for the
 evolution of $f$ between horizon crossing $\eta \approx 1/k_S$, when the curvature fluctuation for $k_S$ froze out, and $\eta_f$, the epoch of slow-roll violation.  
They form a multiplicative correction to $I_{1B}$ and $I_{2B}$ of
\begin{equation}
R \approx \Big\{
\begin{array}{cc} 
 1 + {\bar n_s-1 \over 2} \ln \left( { k_S \eta_f \over x_{\rm max}} \right),  &   k_S\eta_f < x_{\rm max},    \\
 1, &   k_S\eta_f \ge x_{\rm max},  \\
\end{array}\label{eqn:SLSRcorrection}
\end{equation}
with $x_{\rm max}= e^{2-\gamma_E}/2 \approx 2.07$ where $\gamma_E$ is the
Euler-Mascheroni constant.  Note that the $\ln(k\eta_f)$ term is simply the leading order
slow-roll expansion of $\Delta_{\curv}(k_S) / \Delta_{\curv}(\eta_f^{-1})$ 
as one might expect.  Furthermore, the GSR technique can be shown to imply a slow-roll freeze out
at $k\eta = e^{7/3-\gamma_E}/2$ rather than $=1$ (see \cite{Stewart:2001cd}, Eq.~105).
In Fig. \ref{fig:sqlimitlowc} (lower), we demonstrate the effect of this rescaling with Eq.\ (\ref{eqn:SLSRcorrection}), denoted GSR$_{\rm 1S}$.  Here the $k_S$ dependent discrepancy
disappears entirely.

Applying this correction does not impact any triangle where all three modes
are subhorizon at $\eta_f$ and hence has very little impact on high signal-to-noise modes. 
Furthermore given that  $\eta_f$ must be comparable to the horizon size for the bispectrum features to be detectable in the CMB,  observable triangles cannot acquire large logarithmic corrections. We conclude that for practical purposes, this correction can be safely ignored.

\begin{figure}[t]
\centerline{
\psfig{file=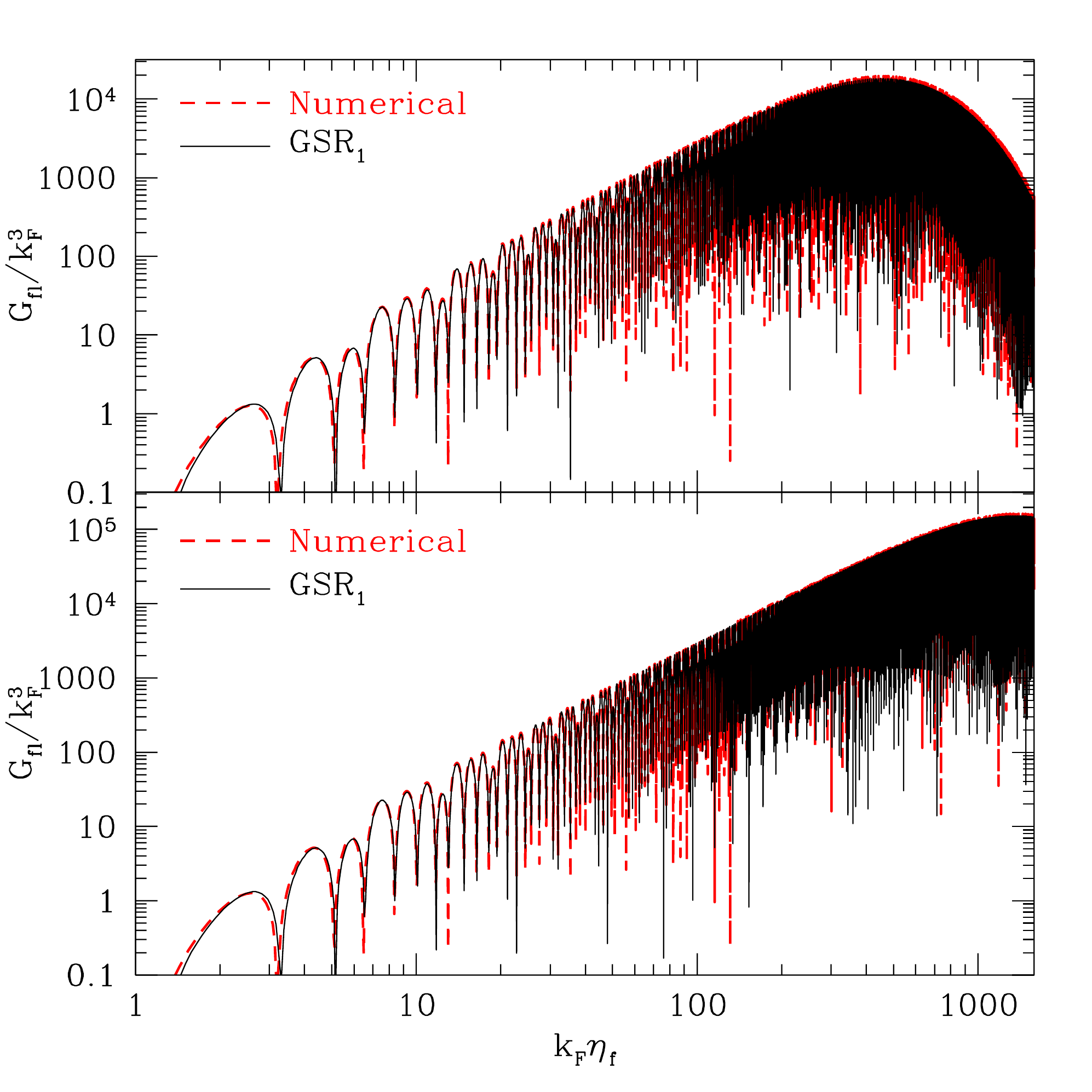, width=3.5in}
}
\caption{
Flat limit bispectrum for a  large amplitude step $c=5.75 \times 10^{-4}$ with $d=0.0003$.   Flat triangle agreement between the GSR$_{1}$ approximation and 
numerical evaluation is comparable to equilateral triangles.}
\label{fig:flat}
\end{figure}

{
At higher values of the step height, $c$, our scaling still removes the $k_S$ dependent
errors  (see Fig.~\ref{fig:sqlimithighc}).   This is because these errors correspond to
the slow evolution of $f$ between $\eta=1/k_S$ and $\eta_f$ rather than the feature
itself.    On the other hand, a new scaling offset develops that can be attributed to the
change the feature makes on $f$ at $\eta_f$.   In principle these could be corrected by
integrating the feature contributions to $G'$ in the first order $U_{1B}$ and $U_{1D}$ terms.
However this offset is of the same order and nature as those found for the equilateral cases.  Given that
squeezed triangles do not dominate the signal-to-noise, we again conclude
that further correction is unnecessary.}

In Fig.~\ref{fig:sqlargec}, 
as a test of the consistency relation itself for a large amplitude step, we compare the numerical power spectrum slope to the squeezed bispectrum. The operator which dominates the bispectrum in the case of a feature is a subleading contribution to the usual slow-roll consistency relation. Thus, we subtract off the slow-roll contribution to the slope of the numerical power spectrum, and instead compare the squeezed limit bispectrum with the deviation away from the slow roll result  for the slope of the power spectrum, $\bar{n}_s - n_{s}(k_{L})$, where $\bar{n}_{s} \approx 0.963$ is the slope of the power spectrum in the absence of the feature.  In practice this correction is negligible if $k_L \eta_f \gg 1$ or more generally when $|f_{NL} |\gg  {\cal O}(1-\bar n_s)$.   It is only that the agreement is so good that we 
take it into account here.   Furthermore, the excellent agreement checks
 the accuracy of the numerical bispectrum calculation.

Finally  it is interesting to note that flat configurations where $k_1 = 2k_2=2k_3 =k_F$ are 
somewhat special in Eq.~(\ref{eqn:gsrlbi}), since the argument to the  $I_5$ integrals  vanishes leaving the expression ill-defined.  It is in fact well-defined if we take instead the limit
of a flat triangle $k_\epsilon = K-2k_F \rightarrow 0$
\begin{eqnarray}
&&
B_{\cal R}(k_F,k_{F}/2,k_{F}/2)  \approx  {(2 \pi)^4  \over k_F^6 } {4 \Delta_{\cal R}(k_F) \Delta_{\cal R}^2(k_F/2) }\nonumber\\&&\qquad 
\Big\{- I_{B0}(2 k_F) - 7 I_{B1}(2 k_F) + 12 I_{B2}(2 k_F) 
\nonumber\\
&& \qquad   +{ I_1(k_F) \over \sqrt 2}  \Big[ - I_{B3}(2 k_F) - 7 I_{B4}(2 k_F) + 12 I_{B5}(2 k_F) 
\nonumber\\
&& \qquad  
- 6 {k_\epsilon \over k_F} I_{B5}(k_\epsilon) \Big]  \label{eqn:flat}\\
&& \qquad   +{ 2I_1(k_F/2) \over \sqrt 2}  \Big[ - I_{B3}(2 k_F) - 7 I_{B4}(2 k_F)+ 2 I_{B4}(k_F) 
\nonumber\\
&& \qquad  
+ 12 I_{B5}(2 k_F) - 6  I_{B5}(k_F)\Big] \Big\},\nonumber
\end{eqnarray}
which again becomes independent of $\eta_*$ as the triangle flattens.
In Fig.~\ref{fig:flat}, we compare our first order approximation to the numerical results. 
For scales between the first oscillation and the damping tail, the approximation captures
the flat and equilateral bispectrum behavior comparably well.

\section{Discussion}
\label{sec:discussion}

We have devised an efficient method to compute first-order corrections to the bispectrum
of inflationary models with features.     First order corrections to the bispectrum are generically at least $\sim$10\% and can reach order unity for order unity features in the power spectrum.

Based on the GSR approach, we have shown that 
corrections in the  high $k$ limit, where the bispectrum amplitude is set well
within the horizon, take on a simple form involving only single integrals over the slow-roll parameters.   This limit is observationally important since it is here that the bispectrum
can become  large if the inflaton crosses a feature in the potential in much less than an e-fold of the expansion.   We have constructed these corrections such that they implement consistency relations at low $k$ in a controlled fashion.

Comparison with direct numerical computation of the bispectrum shows that the approximation works extremely well for the full range where the bispectrum is large.   For
cases where the zeroth order expressions deviate by 50\%, the first order approximation deviates
by less than 10\%.  
These techniques should be useful for analyzing any model where inflaton potential features
 provide observably large non-Gaussianity.

\acknowledgements

This work was supported in part by the Kavli Institute for Cosmological Physics at the University of Chicago through grants NSF PHY-0114422 and NSF PHY-0551142 and an endowment from the Kavli Foundation and its founder Fred Kavli. WH was additionally supported by U.S.~Dept.\ of Energy contract DE-FG02-90ER-40560 and the David and Lucile Packard Foundation. 

\vfill

\bibliography{bifo}

\end{document}